\documentclass[11pt,titlepage,a4paper]{article}
\usepackage{bozhomac} 
\usepackage{bozhlogo}
\usepackage{xr1,cite}
\title{\bfseries \vspace*{-1.678902345in}
\vspace*{-3ex}
{
\begin{flushright}
	\textbf{\large LANL xxx E-print archive No. quant-ph/9902067}\\[2ex]
\end{flushright}
}
\huge Comments on:\\
\LARGE
``Quantum mechanics as a gauge theory of metaplectic spinor fields''
by M. Reuter\\[1ex]
\large
[Int. J. Mod. Phys. vol.~A13 (1998), 3835--3884;\\
LANL E-print No.~hep-th/9804036]
}

\vspace{1.7ex}

\author{
Bozhidar Z. Iliev\;%
\thanks{Department Mathematical Modeling,
Institute for Nuclear Research and \mbox{Nuclear} Energy,
Bulgarian Academy of Sciences,
Boul. Tzarigradsko chauss\'ee~72, 1784 Sofia, Bulgaria}\,%
\thanks{E-mail address: bozho@inrne.bas.bg}\,%
\thanks{URL: http://www.inrne.bas.bg/mathmod/bozhome/}
}

\date{
 \vspace{2.27ex}%
\ShortTitle{Comments on: ``Quantum mechanics as a \ldots'' }\\[0.27ex]
 \vspace{3.27ex}
	\begin{tabular}{r@{$\colon\to~$}l}
 \vspace{0.09ex} Basic ideas	& June 24, 1998	\\[0.09ex]
 \vspace{0.09ex} Began		& July 3, 1998	\\[0.09ex]
 \vspace{0.09ex} Ended		& July 7, 1998	\\[0.09ex]
\vspace{0.09ex} Initial typeset& July 5--7, 1998 \\[0.09ex]
 \vspace{0.09ex} Revised	& November 1--7, 1998	\\[0.09ex]
 \vspace{0.09ex} Last update	& February 19, 1999	\\[0.09ex]
 \vspace{0.27ex} Produced	& \fbox{\today}		\\[0.27ex]
	\end{tabular} \\[1.27ex]
	\begin{tabular}{r@{$\colon~$}l}
\vspace{0.27ex} LANL xxx archive server E-print No.& quant-ph/9902067
							\\[0.27ex]
	\end{tabular} \\[-0.27ex]
 \vspace{4.27ex}{\Huge\BOZHO}	\\[4.27ex]
\vspace{0.27ex}\Subject{Quantum mechanics; Differential geometry} \\[2.27ex]
	\begin{tabular}{r@{\hspace{0.512em}}|@{\hspace{0.512em}}l}
\vspace{0.27ex}		  
&
\vspace{0.27ex}
	\end{tabular} \\[1.27ex]
\vspace{0.27ex}	\\[0.27ex]
}

\newcommand{\Hil}{\mathcal{F}} 
\newcommand{\HilB}{(\bHil,\proj,\base)} 
 \newcommand{\bHil}{\mathit{F}} 
 \newcommand{\proj}{\pi}  
 \newcommand{\base}{\mathit{M}} 

\begin{document}  

\pagestyle{myheadings}
\markright{\itshape\bfseries Bozhidar Z. Iliev:
 	   \upshape\sffamily\bfseries
		Comments on ``Quantum mechanics as a \ldots''}

\renewcommand{\thefootnote}{\fnsymbol{footnote}}
\maketitle			
\renewcommand{\thefootnote}{\arabic{footnote}}

\tableofcontents

\begin{abstract}

	We point out how some mathematically incorrect passages
of~\cite{Reuter} can be formulated in a rigorous way. The fibre bundle
approach to quantum mechanics
of~\cite{bp-BQM-introduction+transport,bp-BQM-equations+observables,
	bp-BQM-pictures+integrals,bp-BQM-mixed_states+curvature,
	bp-BQM-interpretation+discussion}
is compared with the one contained in \emph{loc.\ cit.}

\end{abstract}

\section{Introduction}	\label{Introduction}

	The purpose of these notes is the comparison of fibre bundle
approach to (non-relativistic) quantum mechanics developed in our
investigation~\cite{bp-BQM-introduction+transport,bp-BQM-equations+observables,
	bp-BQM-pictures+integrals,bp-BQM-mixed_states+curvature,
	bp-BQM-interpretation+discussion}
and the one contained in~\cite{Reuter} and the correction of some incorrect
mathematical statements, definitions and expressions in~\cite{Reuter}.
	We have to emphasize that here we shall comment only on some
technical mathematical details of~\cite{Reuter}. What concerns the
`physical' part of this interesting paper, we agree with the author's own
conclusions and will not concern with it in this text. We hope that the
presented here material will help to the improvement of certain not exactly
mathematical rigorous places in~\cite{Reuter}. So, these remarks may be
considered as a mathematical appendix to \emph{loc.\ cit.}\/\ in which are
given more or less complete instructions how this work can be make
mathematically rigorous.

	Here we freely make use of the notation and terminology
of~\cite{bp-BQM-introduction+transport,bp-BQM-equations+observables,
	bp-BQM-pictures+integrals,bp-BQM-mixed_states+curvature,
	bp-BQM-interpretation+discussion}
to which papers the reader is referred for details and explanations.
	The references to sections, equations, footnotes etc.\
from the parts of the
series~\cite{bp-BQM-introduction+transport,bp-BQM-equations+observables,
	bp-BQM-pictures+integrals,bp-BQM-mixed_states+curvature,
	bp-BQM-interpretation+discussion}
are denoted by the corresponding sequential reference numbers in these parts
preceded by the Roman number of the part in which it appears and a dot as a
separator. For instance, Sect.~I.5 and (IV.2.11) mean respectively
section 5 of part~I, i.e.\  of~\cite{bp-BQM-introduction+transport}, and
equation~(2.11) (equation~11 in Sect.~2) of
part~IV, i.e.\  of~\cite{bp-BQM-mixed_states+curvature}.

	At this point we want to say a few words on the possibility to
identify the Hilbert bundle's base%
\footnote{%
Here and below, when talking about a Hilbert bundle we mean the one used for
the fibre bundle description of quantum
mechanics~\cite{bp-BQM-introduction+transport,bp-BQM-equations+observables,
	bp-BQM-pictures+integrals,bp-BQM-mixed_states+curvature,
	bp-BQM-interpretation+discussion}.%
}
$\base$ with the phase space of certain
system which case is taken as a base for a bundle approach to quantum
mechanics in~\cite{Reuter}.  Our generic opinion is that the phase space is
not a `suitable' candidate for a bundle's base, the reason being the
Heisenberg uncertainty principle by virtue of which the points of the phase
space have no physical meaning~\cite[chapter~IV]{Messiah-1}. This reason does
not apply if as a base is taken the phase space of some observer as, by
definition, the observers are treated as classical objects (systems).
Therefore one can set the base $\base$ of the Hilbert bundle $\HilB$ to be
the phase space of some observer.  Then the reference path $\gamma\colon
J\to\base$ can be interpreted as the observer's phase\ndash space trajectory
which, generally, can have self\ndash intersections. The further treatment of
this case is the same as of $\base=\mathbb{E}^3$.
	Regardless of the above\ndash said, one can always identify $\base$
with the system's phase space, if it exists, as actually $\base$ is a free
parameter
in~\cite{bp-BQM-introduction+transport,bp-BQM-equations+observables,
	bp-BQM-pictures+integrals,bp-BQM-mixed_states+curvature,
	bp-BQM-interpretation+discussion}.

	An interesting bundle approach to quantum mechanics
is contained in~\cite{Reuter}. In it the evolution of a quantum system is
described in a Hilbert bundle \emph{over the system's phase space} with the
ordinary system's Hilbert space as a (typical) fibre which is, some times,
identified with the fibre over an arbitrary fixed phase-space point. The
evolution itself is presented as a parallel transport in the bundle space
generated via non\ndash dynamical linear (and symplectic) connection which is
closely related to the symplectic structure of the phase space.
	An important feature of~\cite{Reuter} is that in it the bundle
structure is derived from the physical content of the paper. In this
sense~\cite{Reuter} can be considered as a good motivation for the general
constructions
in~\cite{bp-BQM-introduction+transport,bp-BQM-equations+observables,
	bp-BQM-pictures+integrals,bp-BQM-mixed_states+curvature,
	bp-BQM-interpretation+discussion}.

	Before comparing the mathematical results of~\cite{Reuter} with the
ones of~\cite{bp-BQM-introduction+transport,bp-BQM-equations+observables,
	bp-BQM-pictures+integrals,bp-BQM-mixed_states+curvature,
	bp-BQM-interpretation+discussion}
in Sect.~\ref{Conclusions}, we will pay attention in Sect.~\ref{Critique} on
some incorrect `bundle' expressions in~\cite{Reuter} which, however,
happily do not influence most of the conclusions made on their base.

\section{Critical remarks}	\label{Critique}

	In this section we point to and show possible ways for improving of
a number of mathematically non-rigorous or wrong expressions, assertions, and
definitions in~\cite{Reuter}. Once again we emphasize that all this concerns
only the `bundle' part of the mathematical structure of \emph{loc.\ cit.}\
and does not deal with its physical contents.

	First of all, expressions like%
\footnote{%
See, for instance,
\cite[equations~(3.1), (3.9), (3.14), (3.51), (3.52), (4.40), (4.41),
(5.46), (5.47)]{Reuter}.%
}
 $\pd_a|\psi\rangle_\phi$ and $\od|\psi\rangle_{\phi(s)}/\od s$
(in the notation
of~\cite{bp-BQM-introduction+transport,bp-BQM-equations+observables,
	bp-BQM-pictures+integrals,bp-BQM-mixed_states+curvature,
	bp-BQM-interpretation+discussion}
$\phi\in\base$(=system's phase space) and
$\phi\colon J\to\base$, $\phi\colon s\mapsto \phi(s)$ $s\in J$ respectively
and $|\psi\rangle_\phi\in\bHil_\phi$ is an element in the fibre over $\phi$,
\ie $|\psi\rangle$ is a section of the bundle) are not defined as the
defining them (conventional) limits contain differences of elements of
different fibres which are undefined objects \emph{per.\ ce}. Generally this
situation is the same as outlined at the beginning of Sect.~\ref{V}.
For the same reason the difference
$|\psi\rangle_{\phi+\delta\phi} - |\psi\rangle_{\phi}$
in~\cite[equation~(4.10)]{Reuter} is senseless. Almost the same is the
situation with $\pd_a\mathcal{O}_f$
in~\cite[equations~(4.13) and~(5.9)]{Reuter}
(see also~\cite[equation~(5.21)]{Reuter})
where $\mathcal{O}_f$ is, in the terminology
of~\cite{bp-BQM-introduction+transport,bp-BQM-equations+observables,
	bp-BQM-pictures+integrals,bp-BQM-mixed_states+curvature,
	bp-BQM-interpretation+discussion},
a bundle morphism corresponding
to a dynamical variable whose classical analogue is a classical observable
$f\colon\base\to\mathbb{R}$. Since
\(
\mathcal{O}_f \colon \phi \mapsto \mathcal{O}_f(\phi)
				\colon \bHil_\phi\to\bHil_\phi
\),
the derivative  $\pd_a\mathcal{O}_f|_\phi$ can not be defined as
\(
\lim_{\varepsilon\to0}\frac{1}{\varepsilon}
\bigl( \mathcal{O}_f(\phi+\varepsilon_a) - \mathcal{O}_f(\phi) \bigr)
\)
with $(\varepsilon_a)^b = \varepsilon \delta_{a}^{b}$
and needs  special redefinition. The same arguments are applicable to
 $\pd_a\varepsilon$, $\pd_b\Gamma_a$ and $\pd_a A$
appearing in~\cite[equations~(3.8), (3.9), (3.14), (3.21)]{Reuter}.
	All of these deficiencies can easily be corrected by rewriting the
corresponding equations and definitions in component form, i.e., formally, by
adding to them the required component indices (however see below).  Besides,
when defining the components, \eg of $|\psi\rangle\in\Hil$ ($=\mathcal{V}$
in~\cite{Reuter}), the author improperly transfers the notation from the
\emph{typical fibre} $\Hil$ to  the \emph{fibres} $\bHil_\phi$ \emph{over}
$\base$. For example, if  $\{|x\rangle\}$
(in the notation
of~\cite{bp-BQM-introduction+transport,bp-BQM-equations+observables,
	bp-BQM-pictures+integrals,bp-BQM-mixed_states+curvature,
	bp-BQM-interpretation+discussion}:
$\{f_x\}$ with $x$ in system's configuration space)
is a basis in $\Hil$ and $\{\langle x|\}$ -- in $\Hil^*$,
$\langle x| := (|x\rangle)^*$ (in the notation
of~\cite{bp-BQM-introduction+transport,bp-BQM-equations+observables,
	bp-BQM-pictures+integrals,bp-BQM-mixed_states+curvature,
	bp-BQM-interpretation+discussion}:
$\{f^x\}$, $f^x:=(f_x)^*$),
then the author writes~\cite[equations~(1.16) and~(1.17)]{Reuter}:
 $\psi^x(\phi)=\langle x|\psi\rangle_\phi$ and
 $\chi_x(\phi)=\lindex[\langle \chi| x\rangle]{\phi}{}$
for the components of $|\psi\rangle_\phi\in\bHil_\phi$ and
$\lindex[\langle \chi|]{\phi}{}\in\bHil_{\chi}^{*}$.
This is incorrect by two reasons: (i)  $\{\langle x|\}$ is a basis in $\Hil$,
not in $\bHil_\phi$, so the inner product, e.g.,
$\langle x| \psi\rangle_\phi$ is not defined, and (ii) since  the inner
product (dual pairing) $\langle\cdot|\cdot\rangle$ is defined as a map
 $\langle\cdot | \cdot\rangle\colon\Hil^*\times\Hil\to\mathbb{C}$, it can not
be used (directly) for the definition of the components of
$|\psi\rangle_\phi\in\bHil_\phi$. (The same remark is true
for~\cite[equation~(1.20)]{Reuter} defining the components of a bundle
morphism (family of operators or (1,1)\ndash multispinor field in author's
terminology).) This confusion can be met and further in the text
(see, e.g~\cite[equations~(3.3), (3.10), (4.4), (5.37)]{Reuter}). A lucky
exception of this rule is~\cite[first equation~(4.42)]{Reuter} (regardless
of the fact that its l.h.s.\ is not defined). In fact, in the framework
of~\cite{Reuter}, this equation is the corner-stone for solving the above
problems with undefined scalar products and, at the end, with the components
of the vectors in $\bHil_\phi$, $\phi\in\base$. For this purpose, the only
thing one has to do is to \emph{define} the l.h.s.
of~\cite[first equation~(4.42)]{Reuter} through its r.h.s., \ie this equation
has to be converted into definition.%
\footnote{%
This is possible on any linearly connected subset of $\base$ containing the
fixed basic point $\phi_0$. The so obtained scalar products in the fibres
over this set are path-independent and self-consistent by virtue of the used
in~\cite{Reuter} `Abelian' connection.%
}

	In this way the following six problems find natural solutions:
(i)
	The (typical) fibre $\Hil$ is identified with the fibre
$\bHil_{\phi_0}$
for arbitrarily fixed point $\phi_0$ in the phase\ndash space and the
homeomorphisms $l_\phi\colon\bHil_\phi\to\Hil$ are given through
 $l_{\phi}^{-1}\colon|\psi\rangle_{\phi_0}\mapsto|\psi\rangle_\phi$
via~\cite[equation~(4.41)]{Reuter}.
(ii)
	A fibre inner (scalar) product
$\lfloor\cdot,\cdot\rfloor \colon \bHil^*\times\bHil\to\mathbb{C}$
is rigorously defined on any fibre $\bHil_\phi$, $\phi\in\base$
by~\cite[first equation~(4.42)]{Reuter}.
(iii)
	Choosing a basis $\{|x\rangle_\phi\}$ in $\bHil_\phi$, we define the
components of, e.g, $|\psi\rangle_\phi\in\bHil_\phi$  by
 $\psi^x(\phi) := \lfloor |x\rangle_{\phi}^{*} , |\psi\rangle_\phi \rfloor$
with $|x\rangle_{\phi}^{*}$ being the dual of $|x\rangle_{\phi}$.
(iv)
	Applying steps (ii) and (iv), we can rewrite all equations
of~\cite{Reuter} containing inner products or vectors' (or multispinors')
components in such a way that they obtain rigorous mathematical meaning.
(v)
	The previous point makes  strict the above-pointed solution of the
problems with derivatives like $\pd_a|\psi\rangle_{\phi}$.
(vi)
	The~\cite[second equation~(4.42)]{Reuter}, which includes the
`background-quantum split symmetry', becomes a consequence
of~\cite[equations~(4.40) and~(4.41)]{Reuter}.

	The so-described procedure allows us to take off the above-pointed
problems and to give a strict mathematical sense to the (most of the) results
of~\cite{Reuter}.

	In~\cite{Reuter} nowhere a precise definition is given of what
exactly a Hilbert bundle is (the author talks about Hilbert spaces attached
to the phase space points etc.) regardless of the fact that this concept
appears many times in the paper. So, somewhere at the beginning of this work
must be said that a Hilbert bundle is a collection $\HilB$ of a phase space
$\base$ (in the concrete case), a map $\proj\colon\bHil\to\base$, and
$\bHil=\cup_{\phi\in\base} \bHil_\phi$ where $\bHil_\phi=\proj^{-1}(\phi)$
are Hilbert spaces homeomorphic to the system's conventional Hilbert space
$\Hil$. Besides, for the concrete purposes of~\cite{Reuter}, to this
collection should be added the structure group $G$ of unitary operators
acting on $\Hil$.

	And the last serious problem of~\cite{Reuter} deserving mentioning.
In the paragraph following~\cite[equation~(3.4)]{Reuter} we see a mixing of
the meaning of \emph{active} and \emph{passive} transformations of the fibres
and where they are acting. By definition, passive are the transformations
that change only the vectors' components and are due to changes of the bases,
while the active ones are fibres' diffeomorphisms. The author writes: ``In
all fibres $\mathcal{V}_\phi$ ($=\bHil_\phi$ in the notation
of~\cite{bp-BQM-introduction+transport,bp-BQM-equations+observables,
	bp-BQM-pictures+integrals,bp-BQM-mixed_states+curvature,
	bp-BQM-interpretation+discussion}
- B.I.) we may
perform independent changes of their base by means of gauge transformation
 $U\colon\base\to G$, $\phi\mapsto U(\phi)$'', where he defines $G$ as the
group of all unitary operators on the fibre $\Hil$ ($\mathcal{V}$ in his
notation). Two incorrect things are presented here: First, since
$U(\phi)\colon\Hil\to\Hil$ by definition, the operator $U(\phi)$ can not act
on the fibre $\bHil_\phi$ over $\phi$ as this is a different space.%
\footnote{%
The existence of a homeomorphism between $\Hil$ and $\bHil_\phi$ does not
influence this conclusion; it can only help to define correctly a
representation of $G$ on $\bHil_\phi$.%
}
And second, the operator $U(\phi)\colon\Hil\to\Hil$  changes not only the
bases in $\Hil$ but also all its vectors, \ie it is not a simple change of
the bases (passive transformation) in $\Hil$, but
an active transformation  in $\Hil$. The conclusion is that $U(\phi)$ does
not act on $\bHil_\phi$ at all  and it is  not a simple change of the
base neither in $\Hil$ nor in $\bHil_\phi$. What the author really wants to
do, we hope, is the following. Let $U\colon\phi\mapsto U(\phi)$ (be a bundle
morphism) with  $U(\phi)\colon\bHil_\phi\to\bHil_\phi$ being a unitary
operator on $\bHil_\phi$. (The operator $U(\phi)$ not only changes the
bases in $\bHil_\phi$, it transforms $|\psi\rangle_\phi\in\bHil_\phi$ into
 $|\psi\rangle_\phi^\prime:=U(\phi)|\psi\rangle_\phi$
(see~\cite[equation~(3.5)]{Reuter}).)
	Furthermore, the author claims that under $U(\phi)$ from the
covariance of the connection~\cite[equation~(3.6)]{Reuter}  follows its
transformation law~\cite[equation~(3.7)]{Reuter} (containing the undefined
term $\pd_a U(\phi)$). Two important remarks are in order here.
The defined by~\cite[equation~(3.1)]{Reuter} covariant
derivative $\nabla_a\equiv\nabla_a(\Gamma)$ (connection $\Gamma$) is not
correct due to the involved in it undefined term $\pd_a|\psi\rangle_\phi$, but
this can be repaired as described already. And next, the transformations
$U(\phi)\colon\bHil_\phi\to\bHil_\phi$ do not act on the connection at all,
they leave it unchanged! What the author really does is that he \emph{defines}
by~\cite[equation~(3.6)]{Reuter}%
\footnote{%
The index $a$ of $\nabla$ in the r.h.s.\ of this equation is missing.%
}
a \emph{new} covariant derivative
$\nabla_a(\Gamma^\prime)$ (connection $\Gamma^\prime$) associated to (the
bundle morphism) $U$ and having the natural property
\(
\nabla_a(\Gamma^\prime)|\psi\rangle_{\phi}^{\prime}
 =
U(\phi)\bigl( \nabla_a(\Gamma)|\psi\rangle_{\phi} \bigr)
\).
Explicitly this \emph{new} connection is given
by~\cite[equation~(3.7)]{Reuter} which is equivalent to the mentioned its
property. Another possibility is to consider the connection components
(coefficients) with respect to two fields of local bases whose vectors are
connected via $U$. In this case $\nabla_a$ transforms as a vector (with
respect to the index $a$), this law replaces~\cite[equation~(3.6)]{Reuter},
and the connections' coefficients transform according
to~\cite[equation~(3.7)]{Reuter} provided in it all operators are replaced
with their matrices in the bases mentioned.

	Ending with the critical comments on~\cite{Reuter}, we conclude: most
of the final results and conclusions of this interesting paper are valid
provided the above\ndash pointed (and other minor) corrections are made in
it. Below we shall suppose that this is carefully done. On this base we will
compare~\cite{Reuter}
with~\cite{bp-BQM-introduction+transport,bp-BQM-equations+observables,
	bp-BQM-pictures+integrals,bp-BQM-mixed_states+curvature,
	bp-BQM-interpretation+discussion}.

\section{Conclusions}	\label{Conclusions}

	The main common point between~\cite{Reuter}
and~\cite{bp-BQM-introduction+transport,bp-BQM-equations+observables,
	bp-BQM-pictures+integrals,bp-BQM-mixed_states+curvature,
	bp-BQM-interpretation+discussion}
is the consistent application of the fibre bundle theory to (nonrelativistic)
quantum mechanics. But the implementation of this intention is quite
different: in~\cite{Reuter} we see a description of quantum mechanics in a
new, but `frozen', geometrical background based on a non-dynamical linear
connection deduced from the symplectical structure of the system's phase
space, while in the
series~\cite{bp-BQM-introduction+transport,bp-BQM-equations+observables,
	bp-BQM-pictures+integrals,bp-BQM-mixed_states+curvature,
	bp-BQM-interpretation+discussion}
is used a `dynamical geometry' (linear transport
along paths, which may turn to be a parallel one generated by a linear
connection) whose properties depend on the system's Hamiltonian, \ie on the
physical system under consideration itself.

	The fact that in~\cite{Reuter} the system's phase space is taken as a
base of the used Hilbert bundle is not essential since nothing can prevent us
from making the same choice as, actually, the base is not fixed
in~\cite{bp-BQM-introduction+transport,bp-BQM-equations+observables,
	bp-BQM-pictures+integrals,bp-BQM-mixed_states+curvature,
	bp-BQM-interpretation+discussion}.
In~\cite{Reuter}  is partially considered the dynamics of multispinor fields.
This is an interesting problem, but, since it is not primary related to
conventional quantum mechanics, we think it is out of the scope of our
works~\cite{bp-BQM-introduction+transport,bp-BQM-equations+observables,
	bp-BQM-pictures+integrals,bp-BQM-mixed_states+curvature,
	bp-BQM-interpretation+discussion}.
The methods of its solution are outlined in~\cite{Reuter} and can easily be
incorporated within the bundle quantum mechanics
of~\cite{bp-BQM-introduction+transport,bp-BQM-equations+observables,
	bp-BQM-pictures+integrals,bp-BQM-mixed_states+curvature,
	bp-BQM-interpretation+discussion}.

	The fields of (metaplectic) spinors used in~\cite{Reuter} are simply
sections of the Hilbert bundle, while the ``world-line spinors'' in
of \emph{loc.\ cit.}\ are sections along paths in the terminology
of~\cite{bp-BQM-introduction+transport,bp-BQM-equations+observables,
	bp-BQM-pictures+integrals,bp-BQM-mixed_states+curvature,
	bp-BQM-interpretation+discussion}.
The family of operators
$\mathcal{O}_{\phi^a}$ or
$\mathcal{O}_{f}(\phi)$~\cite[equations~(4.8) and~(4.9)]{Reuter} acting on
$\bHil_\phi$ are actually bundle morphisms.

	A central r\^ole in both works plays the `principle of invariance of
the mean values': the mean values (mathematical expectations) of the
morphisms corresponding to the observables  (dynamical variables) are
independent of the way they are calculate. We have used this assumption many
times
in~\cite{bp-BQM-introduction+transport,bp-BQM-equations+observables,
	bp-BQM-pictures+integrals,bp-BQM-mixed_states+curvature,
	bp-BQM-interpretation+discussion}
(see, e.g., Sections~\ref{VI}, \ref{VII}, and~\ref{XI}, in particular,
equations~\eref{6.1''}, \eref{7.5}, \eref{7.11}, \eref{7.17},
and~\eref{11.17}) without explicitly formulating it as a `principle'. But if
one wants to build axiomatically the bundle quantum mechanics, he will be
forced to include this principle (or an equivalent to it assertion) into the
basic scheme of the theory. In~\cite{Reuter}  `the invariance of the mean
values' is mentioned several times and it is used practically in the form of
the `background-quantum split symmetry' principle, explained
in~\cite[sect.~4]{Reuter}
(see, e.g.,~\cite[equation~(4.18)]{Reuter} and the comments after it).
Its particular realizations are written
as~\cite[equation~(4.17) and second equation~(4.42)]{Reuter}
which are equivalent to it in the corresponding context. A consequence of the
mean-value invariance is the `Abelian' character of the compatible with it
connections, expressed by~\cite[equation~(4.14)]{Reuter}, which is a special
case of our result~\cite[equation~(4.4)]{bp-BQM-preliminary}.
In~\cite{Reuter} the mean values are independent of the point at which they
are determined. In the bundle quantum mechanics
of~\cite{bp-BQM-introduction+transport,bp-BQM-equations+observables,
	bp-BQM-pictures+integrals,bp-BQM-mixed_states+curvature,
	bp-BQM-interpretation+discussion}
this is not generally the case
as different points correspond to different time values (see,
e.g.,~\eref{6.1''}). This difference clearly reflects the dynamical character
of the approach
of~\cite{bp-BQM-introduction+transport,bp-BQM-equations+observables,
	bp-BQM-pictures+integrals,bp-BQM-mixed_states+curvature,
	bp-BQM-interpretation+discussion}
and the `frozen' geometrical one of~\cite{Reuter}. In any
case, the principle we are talking about is so important that without it
the equivalence between the bundle and conventional forms of quantum
mechanics can not be established.

	In both works the quantum evolution is described via appropriate
transport along paths:
In~\cite[see, e.g., equations~(3.54) and~(4.53)]{Reuter} this is an `Abelian'
parallel transport along curves, whose holonomy group is
$U(1)$~\cite[equation~(4.38)]{Reuter}, while in the
investigation~\cite{bp-BQM-introduction+transport,bp-BQM-equations+observables,
	bp-BQM-pictures+integrals,bp-BQM-mixed_states+curvature,
	bp-BQM-interpretation+discussion}
is employed a transport along paths uniquely determined by the Hamiltonian
(see Sect.~\ref{IV}) which, generally, need not to be a parallel translation.

	Now we turn our attention on the bundle equations of motion:
in~\cite{bp-BQM-introduction+transport,bp-BQM-equations+observables,
	bp-BQM-pictures+integrals,bp-BQM-mixed_states+curvature,
	bp-BQM-interpretation+discussion}
we have a \emph{single} bundle Schr\"odinger
equation~\eref{5.13} (see also its matrix version~\eref{5.1}), while
in~\cite[equation~(5.54)]{Reuter} there is an \emph{infinite} number of such
equations, one Schr\"odinger equation in each fibre $\bHil_\phi$ for the
system's state vector $|\psi(t)\rangle_\phi$ at every point $\phi\in\base$.%
\footnote{%
Note that the appearing in~\cite[equations~(4.54)--(4.56)]{Reuter} operator
$\mathcal{O}_H$ is an analogue of our matrix-bundle Hamiltonian
(see Sect.~\ref{V}).%
}
Analogous is the situation with the statistical operator (compare our
equation~\eref{11.17} or~\eref{11.15} with~\cite[equation~(4.56)]{Reuter}).
This drastical difference is due to the \emph{different objects} used to
describe systems states: for the purpose we have used \emph{sections along
paths} (see Sect.~\ref{new-I}), while in~\cite{Reuter} are utilized
\emph{(global) sections} of the bundle defined via~\eref{4.3b}
(cf.~\cite[equation~(4.41)]{Reuter}). Hence, what actually is done
in~\cite{Reuter} is the construction of an isomorphic images of the quantum
mechanics from the fibre $\Hil$ in every fibre $\bHil_\phi$, $\phi\in\base$
(see the comments after~\eref{4.3b}).

	To summarize the comments on part of the mathematical
structures in~\cite{Reuter}: It contains a fibre
bundle description of quantum mechanics. The state vectors are replaced by
(global) sections of a Hilbert bundle with the system's phase space as a base
and their (bundle) evolution is governed through Abelian parallel transport
arising from the symplectical structure of the phase space. Locally, in any
fibre of the bundle, the evolution is presented by a Schr\"odinger
equation, specific for each fibre of the bundle. The work contains a number
of incorrect mathematical constructions which, however, can be corrected so
that the final conclusions remain valid. Some ideas of the paper are near to
the ones
of~\cite{bp-BQM-introduction+transport,bp-BQM-equations+observables,
	bp-BQM-pictures+integrals,bp-BQM-mixed_states+curvature,
	bp-BQM-interpretation+discussion}
but their implementation and development is quite different in these
investigations.

	The style and mathematical language of~\cite{Reuter} are typical for
the high energy and particle physics literature in which the mathematical
`details' we are emphasizing in these notes are ``understood and hardly ever
mentioned explicitly''%
\footnote{%
Quotation from an e-mail massage of M.~Reuter to the author (September 27,
1998).%
}.
In this sense, the present work can be considered as an appendix to
\emph{loc.\ cit.}\ in which is pointed how some its passages can be translated
into a manner suitable for mathematicians or mathematical physicists.

\section*{Acknowledgments}

	This work was partially supported by the National Foundation for
Scientific Research of Bulgaria under Grant No.~F642.

\bibliography{bozhopub,bozhoref}

\begin{thebibliography}{1}

\bibitem{Reuter}
Martin Reuter.
\newblock Quantum mechanics as a gauge theory of metaplectic spinor fields.
\newblock {\em Int. J. Mod. Phys.}, A13:3835--3884, 1998.
\newblock (see also: preprint DESY~97-127, 1997 and LANL xxx archive server,
  E-print No. hep-th/9804036).

\bibitem{bp-BQM-introduction+transport}
Bozhidar~Z. Iliev.
\newblock Fibre bundle formulation of nonrelativistic quantum mechanics. {I}.
  {Introduction}. {The} evolution transport.
\newblock (LANL xxx archive server, E-print No. quant-ph/9803084) Submitted to
  J. Physics A: Math. \& Gen., 1998.

\bibitem{bp-BQM-equations+observables}
Bozhidar~Z. Iliev.
\newblock Fibre bundle formulation of nonrelativistic quantum mechanics. {II}.
  {Equations} of motion and observables.
\newblock (LANL xxx archive server, E-print No. quant-ph/9804062) Submitted to
  J. Physics A: Math. \& Gen., 1998.

\bibitem{bp-BQM-pictures+integrals}
Bozhidar~Z. Iliev.
\newblock Fibre bundle formulation of nonrelativistic quantum mechanics. {III}.
  {Pictures} and integrals of motion.
\newblock (LANL xxx archive server, E-print No. quant-ph/9806046) Submitted to
  J. Physics A: Math. \& Gen., 1998.

\bibitem{bp-BQM-mixed_states+curvature}
Bozhidar~Z. Iliev.
\newblock Fibre bundle formulation of nonrelativistic quantum mechanics. {IV}.
  {Mixed} states and evolution transport's curvature.
\newblock (LANL xxx archive server, E-print No. quant-ph/9901039), 1999.

\bibitem{bp-BQM-interpretation+discussion}
Bozhidar~Z. Iliev.
\newblock Fibre bundle formulation of nonrelativistic quantum mechanics. {V}.
  {Theory's} interpretation, summary, and discussion.
\newblock (LANL xxx archive server, E-print No. quant-ph/9902068), 1998.

\bibitem{Messiah-1}
A.~M.~L. Messiah.
\newblock {\em Quantum mechanics}, volume~I.
\newblock North Holland, Amsterdam, 1961.

\bibitem{bp-BQM-preliminary}
Bozhidar~Z. Iliev.
\newblock Quantum mechanics from a geometric-observer's viewpoint.
\newblock {\em Journal of Physics A: Mathematical and General},
  31(4):1297--1305, January 1998.
\newblock (LANL xxx archive server, E-print No. quant-ph/9803083).

\end{thebibliography}
\bibliographystyle{unsrt}

\end{document}